\documentclass[twocolappendix,appendixfloats,]{emulateapj}

\usepackage[colorlinks = true, linkcolor = blue, urlcolor  = blue, citecolor = blue, anchorcolor = blue]{hyperref}
\usepackage{hyperref}
\usepackage{apjfonts}
\usepackage{rotating}
\usepackage{amsmath}
\usepackage{color}
\usepackage{color}
\usepackage{CJK}
\newcommand{\pivec}{\mbox{\boldmath $\pi$}}
\newcommand{\muvec}{\mbox{\boldmath $\mu$}}
\newcommand{\deltavec}{\mbox{\boldmath $\delta$}}

\newcommand{\thetae}{\theta_{\rm E}}
\newcommand{\pie}{\pi_{\rm E}}
\newcommand{\pien}{\pi_{{\rm E},N}}
\newcommand{\piee}{\pi_{{\rm E},E}}

\definecolor{darkbrown}{RGB}{139,69,19}

\shorttitle{OGLE-2018-BLG-0022}
\shortauthors{Han et al.}

\begin{document}

\title{ OGLE-2018-BLG-0022: First Prediction of an Astrometric Microlensing Signal 
from a Photometric Microlensing Event}
%A Very Rare Case Microlensing Event With A Uniquely Determined Lens Mass}

\author{
% leading author -----------------------------
Cheongho~Han\altaffilmark{002}, Ian~A.~Bond\altaffilmark{003,102}, 
Andrzej~Udalski\altaffilmark{004,100}, 
Sebastiano~Calchi~Novati\altaffilmark{010,103}, 
Andrew~Gould\altaffilmark{001,005,006,101}, 
Valerio~Bozza\altaffilmark{e01,e02},
Yuki~Hirao\altaffilmark{018,102}, 
Arnaud~Cassan\altaffilmark{e03}
\\
(Leading authors),\\
and \\
% KMTNet ---------------------------
Michael~D.~Albrow\altaffilmark{007}, Sun-Ju~Chung\altaffilmark{001,008},  
Kyu-Ha~Hwang\altaffilmark{001}, Chung-Uk~Lee\altaffilmark{001}, Yoon-Hyun~Ryu\altaffilmark{001},
In-Gu~Shin\altaffilmark{009}, Yossi~Shvartzvald\altaffilmark{010,103}, Jennifer~C.~Yee\altaffilmark{009},
Youn~Kil~Jung\altaffilmark{001,101},
Doeon~Kim\altaffilmark{002}, Woong-Tae~Kim\altaffilmark{e04}
% technical staff
Sang-Mok~Cha\altaffilmark{001,012}, Dong-Jin~Kim\altaffilmark{001}, Hyoun-Woo~Kim\altaffilmark{001}, 
Seung-Lee~Kim\altaffilmark{001,008}, Dong-Joo~Lee\altaffilmark{001}, Yongseok~Lee\altaffilmark{001,012}, 
Byeong-Gon~Park\altaffilmark{001,008}, Richard~W.~Pogge\altaffilmark{004}, Weicheng~Zang\altaffilmark{aa1}\\
(The KMTNet Collaboration),\\
% MOA --------------------------------
Fumio~Abe\altaffilmark{013}, Richard~Barry\altaffilmark{014}, David~P.~Bennett\altaffilmark{014,015},           
Aparna~Bhattacharya\altaffilmark{014,015}, Martin~Donachie\altaffilmark{016}, Akihiko~Fukui\altaffilmark{017},               
Yoshitaka~Itow\altaffilmark{013}, Kohei~Kawasaki\altaffilmark{018},              
Iona~Kondo\altaffilmark{018}, Naoki~Koshimoto\altaffilmark{019,020}, Man~Cheung~Alex~Li\altaffilmark{016},    
Yutaka~Matsubara\altaffilmark{013}, Yasushi~Muraki\altaffilmark{013}, Shota~Miyazaki\altaffilmark{018},                
Masayuki~Nagakane\altaffilmark{018}, Cl\'ement~Ranc\altaffilmark{014}, Nicholas~J.~Rattenbury\altaffilmark{016}, 
Haruno~Suematsu\altaffilmark{018}, Denis~J.~Sullivan\altaffilmark{021}, Takahiro~Sumi\altaffilmark{018},                
Daisuke~Suzuki\altaffilmark{022}, Paul~J.~Tristram\altaffilmark{023}, Atsunori~Yonehara\altaffilmark{024}\\                    
(The MOA Collaboration),\\
% OGLE -----------------------------
Przemek~Mr{\'o}z\altaffilmark{004},
Micha{\l}~K.~Szyma{\'n}ski\altaffilmark{004},
Jan~Skowron\altaffilmark{004},
Radek~Poleski\altaffilmark{005},
Igor~Soszy{\'n}ski\altaffilmark{004},
Pawe{\l}~Pietrukowicz\altaffilmark{004},
Szymon~Koz{\l}owski\altaffilmark{004},
Krzysztof~Ulaczyk\altaffilmark{025},
Krzysztof~A.~Rybicki\altaffilmark{004},
Patryk~Iwanek\altaffilmark{004},
Marcin~Wrona\altaffilmark{004}\\
(The OGLE Collaboration) \\   
% Spitzer -----------------------------
Charles~A.~Beichman\altaffilmark{010}, Geoffery~Bryden\altaffilmark{s01}, Sean~Carey\altaffilmark{010},
B.~Scott~Gaudi\altaffilmark{005}, Calen~B.~Henderson\altaffilmark{010}\\
({\it Spitzer} Microlensing Team) \
}

\email{cheongho@astroph.chungbuk.ac.kr}

%%=====================================
\altaffiltext{001}{Korea Astronomy and Space Science Institute, Daejon 34055, Republic of Korea} 
\altaffiltext{002}{Department of Physics, Chungbuk National University, Cheongju 28644, Republic of Korea} 
\altaffiltext{003}{Institute of Natural and Mathematical Sciences, Massey University, Auckland 0745, New Zealand}
\altaffiltext{004}{Warsaw University Observatory, Al.~Ujazdowskie 4, 00-478 Warszawa, Poland} 
\altaffiltext{005}{Department of Astronomy, Ohio State University, 140 W. 18th Ave., Columbus, OH 43210, USA} 
\altaffiltext{006}{Max Planck Institute for Astronomy, K\"onigstuhl 17, D-69117 Heidelberg, Germany} 
\altaffiltext{007}{University of Canterbury, Department of Physics and Astronomy, Private Bag 4800, Christchurch 8020, New Zealand} 
\altaffiltext{008}{Korea University of Science and Technology, 217 Gajeong-ro, Yuseong-gu, Daejeon, 34113, Republic of Korea} 
\altaffiltext{009}{Harvard-Smithsonian Center for Astrophysics, 60 Garden St., Cambridge, MA 02138, USA} 
\altaffiltext{010}{IPAC, Mail Code 100-22, Caltech, 1200 E.\ California Blvd., Pasadena, CA 91125, USA}
\altaffiltext{012}{School of Space Research, Kyung Hee University, Yongin, Kyeonggi 17104, Korea} 
\altaffiltext{aa1}{Physics Department and Tsinghua Centre for Astrophysics, Tsinghua University, Beijing 100084, China} 
% -----------------
\altaffiltext{013}{Institute for Space-Earth Environmental Research, Nagoya University, Nagoya 464-8601, Japan}
\altaffiltext{014}{Code 667, NASA Goddard Space Flight Center, Greenbelt, MD 20771, USA}
\altaffiltext{015}{Department of Astronomy, University of Maryland, College Park, MD 20742, USA}
\altaffiltext{016}{Department of Physics, University of Auckland, Private Bag 92019, Auckland, New Zealand}
\altaffiltext{017}{Okayama Astrophysical Observatory, National Astronomical Observatory of Japan, 3037-5 Honjo, Kamogata, Asakuchi, Okayama 719-0232, Japan}
\altaffiltext{018}{Department of Earth and Space Science, Graduate School of Science, Osaka University, Toyonaka, Osaka 560-0043, Japan}
\altaffiltext{019}{Department of Astronomy, Graduate School of Science, The University of Tokyo, 7-3-1 Hongo, Bunkyo-ku, Tokyo 113-0033, Japan}
\altaffiltext{020}{National Astronomical Observatory of Japan, 2-21-1 Osawa, Mitaka, Tokyo 181-8588, Japan}
\altaffiltext{021}{School of Chemical and Physical Sciences, Victoria University, Wellington, New Zealand}
\altaffiltext{022}{Institute of Space and Astronautical Science, Japan Aerospace Exploration Agency, 3-1-1 Yoshinodai, Chuo, Sagamihara, Kanagawa, 252-5210, Japan}
\altaffiltext{023}{University of Canterbury Mt.\ John Observatory, P.O. Box 56, Lake Tekapo 8770, New Zealand}
\altaffiltext{024}{Department of Physics, Faculty of Science, Kyoto Sangyo University, 603-8555 Kyoto, Japan}
\altaffiltext{025}{Department of Physics, University of Warwick, Gibbet Hill Road, Coventry, CV4 7AL, UK} 
% -----------------
\altaffiltext{e01}{Dipartimento di Fisica ``E.~R.~Caianiello'', Universit\'e di Salerno, Via Giovanni Paolo II, I-84084 Fisciano (SA), Italy}
\altaffiltext{e02}{Istituto Nazionale di Fisica Nucleare, Sezione di Napoli, Via Cintia, I-80126 Napoli, Italy}
\altaffiltext{e03}{Sorbonne Universit\'es, UPMC Univ Paris 6 et CNRS, UMR 7095, Institut d'Astrophysique de Paris, 98 bis bd Arago, F-75014 Paris, France}
\altaffiltext{e04}{Department of Physics \& Astronomy, Seoul National University, Seoul 151-742, Republic of Korea}
% -----------------
\altaffiltext{s01}{Jet Propulsion Laboratory, California Institute of Technology, 4800 Oak Grove Drive, Pasadena, CA 91109, USA}
%=====================================
\altaffiltext{100}{OGLE Collaboration.}
\altaffiltext{101}{KMTNet Collaboration.}
\altaffiltext{102}{MOA Collaboration.}
\altaffiltext{103}{{\it Spitzer} Microlensing Team.}

\begin{abstract}
In this work, we present the analysis of the binary microlensing event OGLE-2018-BLG-0022
that is detected toward the Galactic bulge field.  The dense and continuous coverage with 
the high-quality photometry data from ground-based observations combined with the space-based 
{\it Spitzer} observations of this long time-scale event enables us to uniquely determine 
the masses $M_1=0.40 \pm 0.05~M_\odot$ and $M_2=0.13\pm 0.01~M_\odot$ of the individual 
lens components.  Because the lens-source relative parallax and the vector lens-source 
relative proper motion are unambiguously determined, we can likewise unambiguously predict 
the astrometric offset between the light centroid of the magnified images (as observed by 
the {\it Gaia} satellite) and the true position of the source.  This prediction can be 
tested when the individual-epoch {\it Gaia} astrometric measurements are released.
\end{abstract}


\section{Introduction}\label{sec:one}

Full characterization of a microlens 
requires the determination of
its mass, $M$. For this 
determination, one has to measure both the angular Einstein radius, $\thetae$, and the
microlens parallax, $\pie$, i.e.,
\begin{equation}
M={\thetae\over \kappa\pie},
\label{eq1}
\end{equation}
where $\kappa= 4G/(c^2~{\rm au})$. In order to determine $\thetae$, one has to detect 
deviations in the lensing light curve caused by finite source effects. For single-lens 
events, which comprise the overwhelming majority of lensing events, it is difficult to 
measure $\thetae$ because finite-source effects can be detected only for the very rare 
cases in which the lens passes over the surface of the source star. Determination of 
$\pie$ requires to detect subtle deviations in the lensing light curve caused by the 
positional change of the source induced by the orbital motion of Earth around the Sun 
\citep{Gould1992}. However, such a parallax signal can be securely detected only for 
long time-scale events observed with a high-quality photometry. As a result, lens mass 
determinations by simultaneously measuring $\thetae$ and $\pie$ have been confined to 
a very small fraction of all events.

In addition to photometric data, lens characteristics can be further constrained with 
astrometric data. By measuring the lensing-induced positional shifts of the 
image centroid using a high-precision instrument
such as {\it Gaia} \citep{Gaia2016}, the 
chance to measure the lens mass becomes higher \citep{Paczynski1996}. Astrometric 
microlensing observation is especially important to detect black holes (BHs), 
including isolated BHs, because the time scales are long for events produced by BHs 
and thus the probability to measure $\pie$ is high. One can also measure the angular 
Einstein radius because astrometric lensing signals scale with $\thetae$.

To empirically verify the usefulness of {\it Gaia} lensing observations in characterizing
microlenses, it is essential to predict {\it Gaia} astrometric signature on real microlensing
events, and then confirm that the actual observations agree with these 
predictions.\footnote{There have been predictions of astrometric microlensing detections 
in the era of {\it Gaia} and other space missions. These have been done by investigating 
catalogs of nearby stars with large proper motions that will approach close to background 
stars, e.g., \citet{Salim2000}, \citet{Proft2011}, and \citet{Sahu2014}.  
\citet{Bramich2018a} and \citet{Bramich2018b} also made predictions of 
astrometric microlensing events using Gaia data release 2.  However, there has been 
no prediction of astrometric lensing signals based on the analysis of photometric 
lensing data.} To do this, one needs a lensing event with not only a robust and unique 
solution but also relatively bright source. Unfortunately, such events are very rare 
because lensing solutions are, in many cases, subject to various types of degeneracy, 
e.g., the close/wide binary degeneracy \citep{Griest1998, Dominik1999, An2005}, the 
ecliptic degeneracy \citep{Smith2003, Skowron2011}, the degeneracy between microlens-parallax 
and lens-orbital effects \citep{Batista2011, Skowron2011, Han2016}, the 4-fold degeneracy 
for space-based microlens parallax measurement \citep{Refsdal1966, Gould1994, Zhu2015}, etc.

In this work, we present the analysis of the binary-lens event OGLE-2018-BLG-0022.  
The dense and continuous coverage with high-quality photometric data from  ground-based 
observations combined with space-based 
{\it Spitzer} observations of this long-timescale 
event yields a lensing solution without any degeneracy, leading to the unique determination 
of the physical lens parameters.  Combined with the relative bright source, the event is 
an ideal case in which astrometric lensing signals can be predicted from the photometric 
data and confirmed from astrometric observations.

\section{Observation and Data}\label{sec:two}

% Figure 1 ------------------------------------------------------
\begin{figure}
\includegraphics[width=\columnwidth]{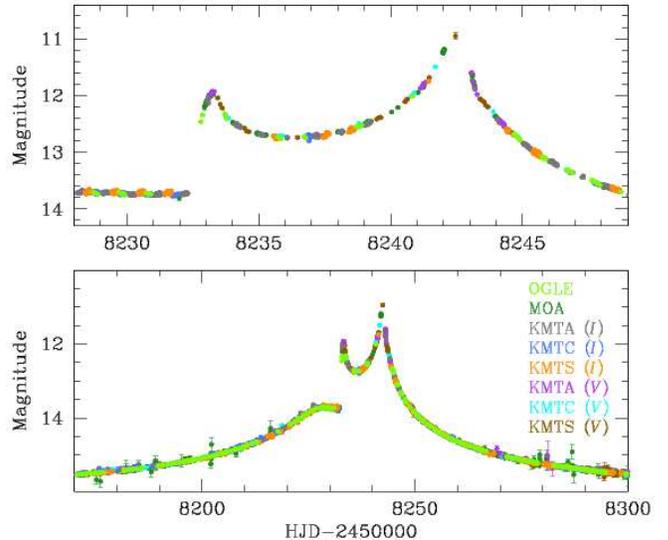}
%\epsscale{0.75}
%\plotone{lc.eps}
\caption{
Light curve of OGLE-2018-BLG-0022. The upper panels shows the zoom of the peak region. 
The colors of the data points match those of the labels of the telescopes used for 
observations.
\vskip0.3cm
}
\label{fig:one}
\end{figure}
% --------------------------------------------------------------

The microlensing event OGLE-2018-BLG-0022 occurred on a star located toward the
Galactic bulge field. The equatorial and galactic coordinates of the source are 
$({\rm RA}, {\rm decl.})_{\rm J2000}=(17:59:27.04, -28:36:37.0)$ and 
$(l, b)=(1.82^\circ, -2.44^\circ)$, respectively. The source had 
a bright baseline magnitude of $I_{\rm base}=15.6$.

% Figure 2 ------------------------------------------------------
\begin{figure*}
\epsscale{0.80}
\plotone{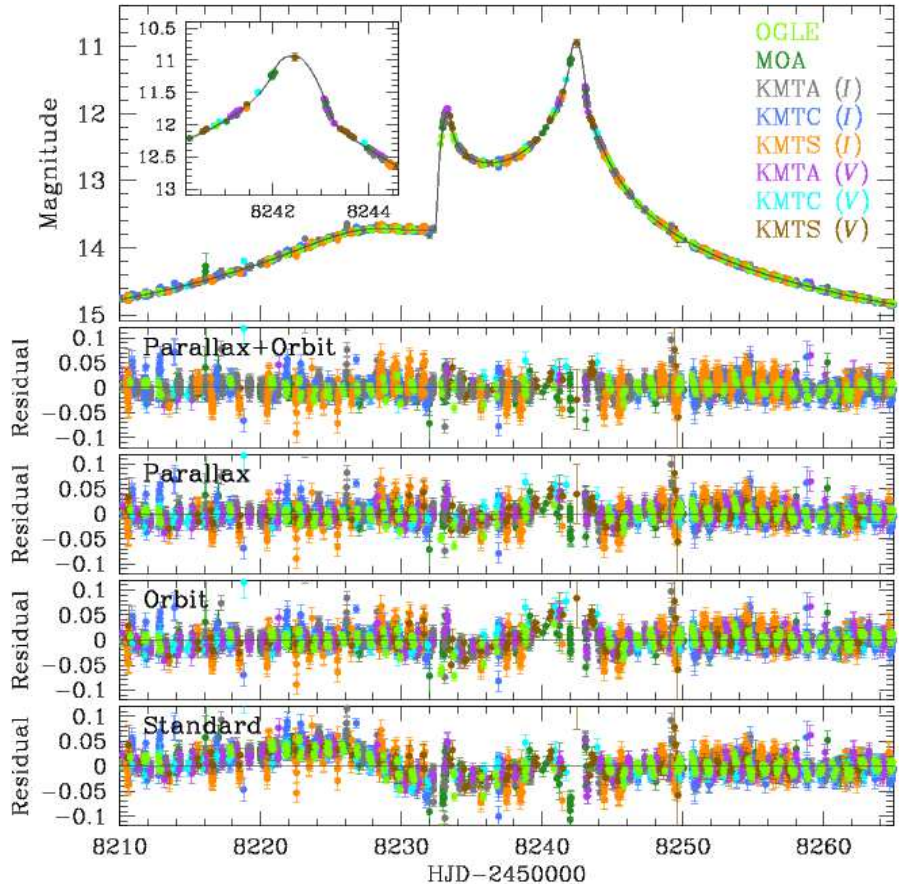}
\caption{
Comparison of models obtained considering various higher-order effects. The top 
panel shows the data near the peak of the light curve, and the lower panels show 
the residuals from the tested models. The model curve superposed on the data points 
is the best-fit model considering both the microlens-parallax and lens-orbital effects.
\vskip0.5cm
}
\label{fig:two}
\end{figure*}
% --------------------------------------------------------------

Figure~\ref{fig:one} shows the light curve of the event. It is characterized by two 
caustic-crossing spikes that occurred at ${\rm HJD}^\prime\equiv {\rm HJD}-2450000\sim 8233.0$ 
and $\sim 8242.5$ and a hump centered at ${\rm HJD}^\prime\sim 8227$.  Caustics produced 
by a binary lens form close curves, and thus caustic crossings occur in multiples of two. 
Combined with the characteristic ``U''-shape magnification pattern between the caustic 
spikes, the first and second spikes are inferred to be produced by the source's caustic 
entrance and exit, respectively. The magnification pattern during the caustic entrance 
exhibits the typical shape when the source passes a regular fold caustic \citep{Schneider1986}.  
However, the pattern during the second caustic spike appears to be different from a regular 
one, suggesting that another caustic feature is involved.  The hump is likely to be produced 
by the source approach to a cusp of a caustic.

The event was discovered in the early 2018 bulge season by the OGLE survey 
\citep{Udalski2015}. The event is registered in the `OGLE-IV Early Warning System' 
page\footnote{http://ogle.astrouw.edu.pl/ogle4/ews/ews.html} as two identification (ID) 
numbers, OGLE-2018-BLG-0022 and OGLE-2018-BLG-0052. We use the former ID.  The source 
was located also in the fields toward which two other lensing surveys of MOA \citep{Bond2001} 
and KMTNet \citep{Kim2016} were monitoring. In the list of `2018 MOA transient 
alerts'\footnote{http://www.massey.ac.nz/~iabond/moa/alert2018/alert.php}, the event was 
registered as MOA-2018-BLG-031. The lensing-induced brightening of the source started 
during the $\sim 3$-month time gap between the 2017 and 2018 bulge seasons.  During 
this period, the Sun passed the bulge field and thus the event could not be observed.  
For this reason, at the first observation conducted in the 2018 bulge season on 
${\rm HJD}^\prime=8151$ (February 1) the light curve was already $\sim 0.2$ mag brighter 
than the baseline magnitude. After being detected, the event lasted throughout the 2018 
season. When the analysis of the event was completed, we learned that the event was 
additionally observed by the ROME/REA survey\footnote{https://robonet.lco.global/}, 
which is a new survey commenced the 2018 season, and an independent analysis was in 
progress.  We, therefore, conduct analysis based on the OGLE+MOA+KMTNet data sets.

We note that the event was very densely and continuously covered with an excellent
photometric quality. The superb coverage was possible thanks to the high cadence of 
the survey observations conducted using globally distributed telescopes. The OGLE 
and MOA surveys utilize the 1.3~m telescope of the Las Campanas Observatory in Chile 
and the 1.8~m telescope located at Mt.~John Observatory in New Zealand, respectively. 
The KMTNet survey uses 3 identical 1.6~m telescopes located at the Cerro Tololo 
Interamerican Observatory in Chile, the South African Astronomical Observatory, 
South Africa, and the Siding Spring Observatory, Australia. We designate the 
individual KMTNet telescopes as KMTC, KMTS, and KMTA, respectively. The OGLE 
survey observed the event with a cadence of $\sim 5$--6/night, and the KMTNet 
$I$-band and MOA cadences were 15 min and 10 min, respectively. In addition, KMTNet 
observed in $V$ band with a cadence of $\sim 2.5$ hours.  The high photometric 
quality was achieved because the source star was very bright. The event brightness 
near the peak was brighter than $I\sim 12$, and many KMTNet data points near and 
above this limit were saturated.  We exclude these data points. However, the KMTNet 
$V$-band points are not saturated, which is the reason for including them in the 
analysis.  We note that there exist additional data obtained from space-based 
{\it Spitzer} observations.  We will discuss the {\it Spitzer} data and the 
analysis of these data in Section~\ref{sec:three-two}.

The event was analyzed nearly in real time with the progress of the event. With the
detection of the anomaly by the ARTEMiS system \citep{Dominik2008}, the first model
was circulated to the microlensing community by V.~Bozza. A.~Cassan and Y.~Hirao also
circulated subsequent models. As the event proceeded, the models were further refined.

Reduction of the data was carried out using the photometry codes developed by the
individual survey groups: \citet{Udalski2003}, \citet{Bond2001}, and \citet{Albrow2009} 
for the OGLE, MOA, and KMTNet data sets, respectively. All of these codes are based on 
the difference imaging method \citep{Alard1998}. For the KMTC $I$ and $V$-band data sets, 
additional photometry is conducted using the pyDIA code \citep{Albrow2017} to measure 
the source color.

\section{Analysis}\label{sec:three}

\subsection{Modeling Light Curve}\label{sec:three-one}

Considering the characteristic caustic-crossing features, we conduct binary-lens 
modeling of the observed light curve. In the first-round modeling, we assume that 
the observer and the lens components do not experience any acceleration, and thus 
the relative lens-source motion is rectilinear. We refer to this model as the 
``standard'' model.

Standard modeling requires seven lensing parameters, including the time of the 
closest lens-source approach, $t_0$, the separation at that time, $u_0$, the 
event timescale, $t_{\rm E}$, the projected separation, $s$, and the mass ratio 
between the lens components, $q$, the source trajectory angle, $\alpha$, and the 
normalized source radius, $\rho$.  The lengths of $s$, $u_0$, and $\rho$ are 
normalized to $\thetae$.  We compute finite-source magnifications considering 
the limb-darkening variation of the source star's surface brightness.  The profile 
of the surface-brightness is modeled by $S\propto 1-\Gamma_\lambda(1-1.5\cos\psi)$, 
where $\Gamma_\lambda$ denotes the linear limb-darkening coefficient and $\psi$ 
represents the angle between the normal to the source surface and the line of sight 
toward the source center.  We adopt the limb-darkening coefficients from 
\citet{Claret2000} considering the source type.  The determination of the source 
type is discussed in section~\ref{sec:three-three}.  The adopted limb-darkening 
coefficients are $\Gamma_V=0.726$, $\Gamma_R=0.639$, and $\Gamma_I=0.527$.  We 
set the center of mass of the binary lens as the reference position. In the 
preliminary modeling, we conduct a grid search for $s$ and $q$ while the other 
parameters are searched for using a downhill approach based on the Markov chain 
Monte Carlo (MCMC) method. We then refine the solutions found from the preliminary 
search by allowing all parameters to vary.

% Table 1 ------------------------------------------------
\begin{deluxetable}{lcc}
\tablecaption{Comparison of models\label{table:one}}
\tablewidth{240pt}
%\tabletypesize{\small}
\tablehead{
\multicolumn{2}{c}{Model}       &
\multicolumn{1}{c}{$\chi^2$}
}
\startdata
Standard          &              &  21219.8   \\
Orbit             &              &  14539.2   \\
Parallax          & ($u_0>0$)    &  13765.0   \\
--                & ($u_0<0$)    &  13764.5  \\
Orbit + Parallax  & ($u_0>0$)    &  13029.7   \\
--                & ($u_0<0$)    &  13256.0
\enddata
%\tablecomments{ }
\end{deluxetable}
% -------------------------------------------------------------

The standard modeling yields a unique solution with binary lens parameters of 
$(s, q)\sim (0.51, 0.35)$. This indicates that the lens is a binary composed of masses 
of a same order with a projected separation smaller than the angular Einstein radius, 
i.e., close binary ($s<1.0$). The event timescale is $t_{\rm E}\sim 67.6$ days, which 
is substantially longer than typical galactic lensing events.  We check for the possible 
existence of a binary-lens solution with $s>1.0$, wide binary, caused by the close/wide 
binary degeneracy.  We find that the fit of the best-fit wide-binary solution is worse 
than the fit of the close-binary solution by $\Delta\chi^2 > 20000$, indicating that 
the close/wide degeneracy is clearly resolved.

Although the standard model provides a fit that describes the overall light curve, 
we find that the model leaves systematic residuals. This can be seen in the bottom 
panel of Figure~\ref{fig:two}, which shows a relatively small ($\lesssim 0.05$ mag) 
but easily noticed deviation from the standard model in the region around the main 
anomaly features. We also find that the deviation persists throughout the light curve. 
This suggests the need to consider higher-order effects.

Noticing the residual from the standard model, we conduct additional modeling
considering higher-order effects. It is known that two higher-order effects cause
long-term deviations in lensing light curves. The first is the microlens-parallax effect,
which is caused by the acceleration of the observer's motion induced by the orbital
motion of Earth around the Sun \citep{Gould1992}. The other is the lens-orbital effect, which
is caused by the acceleration of the lens motion induced by the orbital motion of the lens
\citep{Dominik1998, Ioka1999}. We test these effects by conducting three sets of additional
modeling. In the ``parallax'' and ``orbit'' models, we separately consider the
microlens-parallax and lens-orbital effects, respectively. In the ``orbit+parallax'' model, we
simultaneously consider both effects. For events affected by the microlens-parallax effect,
there usually exist a pair of degenerate solutions with $u_0>0$ and $u_0<0$. This so-called
`ecliptic degeneracy' is caused by the mirror symmetry of the source trajectory with
respect to the binary axis \citep{Smith2003, Skowron2011}. We check this
degeneracy when the microlens-parallax effect is considered in modeling.

Considering the higher-order effects requires one to include additional parameters in modeling. 
Parallax effects are described by 2 parameters of $\pien$ and $\piee$, which denote the two
components of the microlens-parallax vector, $\pivec_{\rm E}$, projected onto the sky along 
the north and east directions in the equatorial coordinates, respectively. Under the approximation
that the change of the lens position caused by the orbital motion is small, lens-orbital
effects are described by 2 parameters of $ds/dt$ and $d\alpha/dt$, which represent the change
rates of the binary separation and the orientation angle, respectively.

% Figure 3 ------------------------------------------------------
\begin{figure}
\includegraphics[width=\columnwidth]{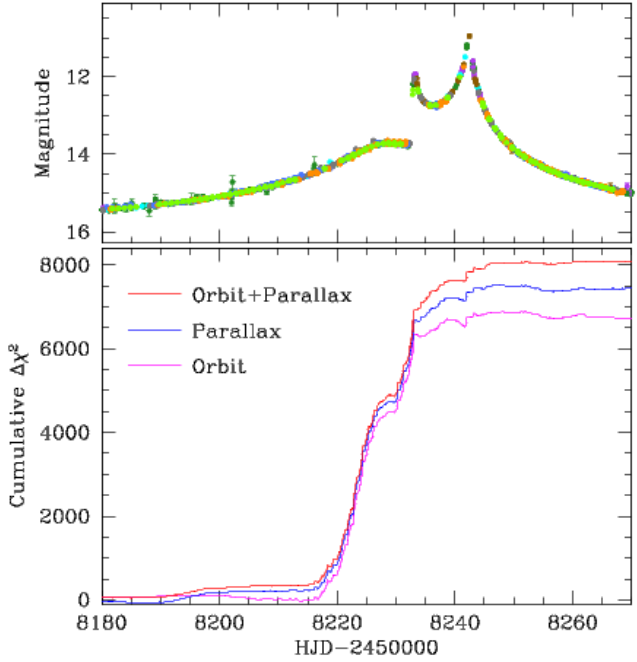}
%\epsscale{0.75}
%\plotone{lc.eps}
\caption{
Cumulative distribution of $\Delta\chi^2$ between the models
with higher-order effects and the standard model.
%\vskip0.5cm
}
\label{fig:three}
\end{figure}
% --------------------------------------------------------------

In Table~\ref{table:one}, we summarize the results of the additional modeling 
runs in terms of $\chi^2$ values.  From the comparison of the model fits, we 
find the following results. First, the fit greatly improves with the consideration 
of the higher-order effects. We find that the fit improves by $\Delta\chi^2= 6680.6$ 
and $7455.3$ with respect to the standard model by considering the lens-orbital and 
microlens-parallax effects, respectively, indicating that the higher-order effects 
are clearly detected. When both effects are simultaneously considered, the fit 
further improves by $\Delta\chi^2\sim 1509.5$ and 734.8 with respect to the orbit 
and parallax models, respectively. To visualize this improvement, we present 
the residuals of the tested models in the lower panels of Figure~\ref{fig:two}. 
In Figure~\ref{fig:three}, we also present the cumulative distributions of 
$\Delta\chi^2$ as a function of time to show the region of the fit improvement. 
It is found that the greatest fit improvement occurs in the region around the 
main features of the light curve, i.e., the hump and caustic spikes, although 
the fit improves throughout the event. Second, the ecliptic degeneracy between 
the solutions with $u_0>0$ and $u_0<0$ is also resolved. It is known that this 
degeneracy is usually very severe even for binary-lens events with well covered 
caustic features, e.g., $\Delta\chi^2\sim 3$ for OGLE-2017-BLG-053 \citep{Jung2018} 
and $\Delta\chi^2\sim 8$ for OGLE-2017-BLG-0039 \citep{Han2018b}. For OGLE-2018-BLG-0022, 
we find that the solution with $u_0>0$ is preferred over the solution with $u_0<0$ 
by $\Delta\chi^2\sim 226.3$, which is big enough to clearly resolve the degeneracy.

% Table 2 ------------------------------------------------
\begin{deluxetable}{lcc}
\tablecaption{Best-fit lensing parameters\label{table:two}}
\tablewidth{240pt}
%\tabletypesize{\small}
\tablehead{
\multicolumn{1}{c}{parameter}            &
\multicolumn{1}{c}{Ground only}          &
\multicolumn{1}{c}{Ground+{\it Spitzer}}
}
\startdata                                              
$t_0$ (${\rm HJD}^\prime$) &  8238.467 $\pm$  0.016   &  8238.490 $\pm$  0.015  \\
$u_0$                      &  0.0085   $\pm$  0.0001  &  0.0084   $\pm$  0.0001 \\
$t_{\rm E}$ (days)         &  71.19    $\pm$  0.29    &  70.41    $\pm$  0.25   \\
$s$                        &  0.528    $\pm$  0.001   &  0.529    $\pm$  0.001  \\
$q$                        &  0.302    $\pm$  0.003   &  0.304    $\pm$  0.003  \\
$\alpha$ (rad)             &  0.176    $\pm$  0.001   &  0.176    $\pm$  0.001  \\
$\rho$ ($10^{-3}$)         &  4.88     $\pm$  0.05    &  4.97     $\pm$  0.04   \\
$\pi_{{\rm E},N}$          &  0.307    $\pm$  0.020   &  0.242    $\pm$  0.020  \\
$\pi_{{\rm E},E}$          &  0.056    $\pm$  0.001   &  0.052    $\pm$  0.001  \\
$ds/dt$ (yr$^{-1}$)        &  0.511    $\pm$  0.020   &  0.443    $\pm$  0.020  \\
$d\alpha/dt$ (yr$^{-1}$)   &  0.506    $\pm$  0.063   &  0.680    $\pm$  0.061  \\
$I_{s,{\rm OGLE}}$         &  15.86    $\pm$  0.005   &  15.86    $\pm$  0.005  \\   
$I_{b,{\rm OGLE}}$         &  18.85    $\pm$  0.070   &  18.85    $\pm$  0.070                      
\enddata                            
\tablecomments{${\rm HJD}^\prime={\rm HJD-2450000}$.
The values $I_{s,{\rm OGLE}}$ and $I_{b,{\rm OGLE}}$ represent the
$I$-band magnitudes of the source and blend estimated based on the OGLE data, 
respectively.
}
\end{deluxetable}
% -------------------------------------------------------------

% Figure 4 ------------------------------------------------------
\begin{figure}
\includegraphics[width=\columnwidth]{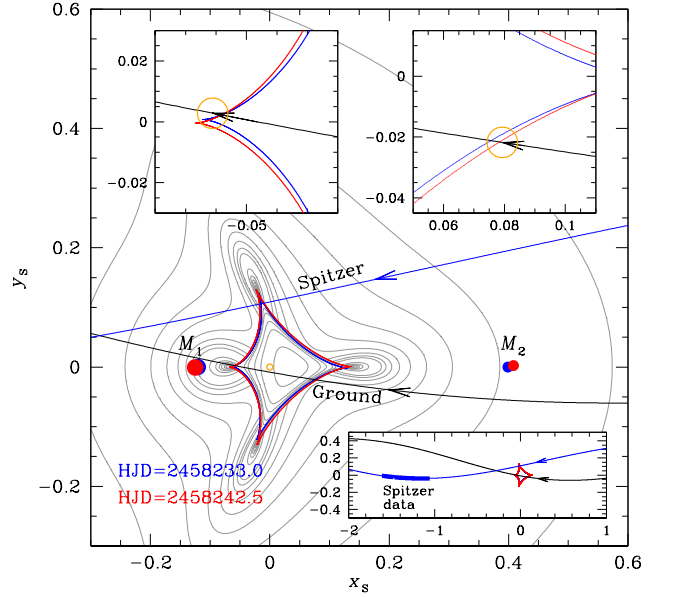}
%\epsscale{0.80}
%\plotone{geo.eps}
\caption{
Configuration of the lens system showing the source motion with respect to the 
caustic.  The black and blue curves with arrows represent the source trajectories 
seen from the ground and the {\it Spitzer} telescope, respectively.  The upper 
right and left insets show zoom of the regions around which the source entered and 
exited the caustic, respectively.  The small orange circle on the source trajectory 
represents the source size.  The grey curves around the caustic represent 
equi-magnification contours. Due to the change of the lens position caused 
by the orbital effect, we present the positions of the lens components ($M_1$ and 
$M_2$) and caustic at two epochs marked inside the main panel.
The lower right inset shows a wider view to present the source positions
when {\it Spitzer} observations were conducted (the region marked by thick line weight).
%\vskip0.5cm
}
\label{fig:four}
\end{figure}
% --------------------------------------------------------------

In the middle column of 
Table~\ref{table:two}, we present the determined lensing parameters of the best-fit 
solution, i.e., orbit+parallax model with $u_0>0$. In Figure~\ref{fig:four}, we also present 
the lens-system configuration, which shows the source trajectory (the black curve with an arrow) 
with respect to the lens components (marked by $M_1$ and $M_2$) and caustic (closed curve 
composed of concave curves). It is found that the source passed almost parallel to the 
binary axis. The caustic, which is composed of 4 folds, is located between the lens 
components. The hump centered at ${\rm HJD}^\prime\sim 8227$ was produced when the 
source passed the excess magnification region extending from the right on-binary-axis 
caustic cusp. The source crossed the lower right fold caustic, producing the first 
caustic spike. Then, the source passed the upper left fold caustic, producing the second 
spike. To be mentioned is that the source enveloped the left on-axis caustic cusp during 
the caustic exit.  See the left inset of Figure~\ref{fig:four}, which shows the enlargement 
of the caustic exit region.  As a result, the caustic-crossing pattern differs from that 
produced when the source passes a regular fold caustic.  See the inset in the upper panel 
of Figure~\ref{fig:two}.

% Figure 5 ------------------------------------------------------
\begin{figure}
\includegraphics[width=\columnwidth]{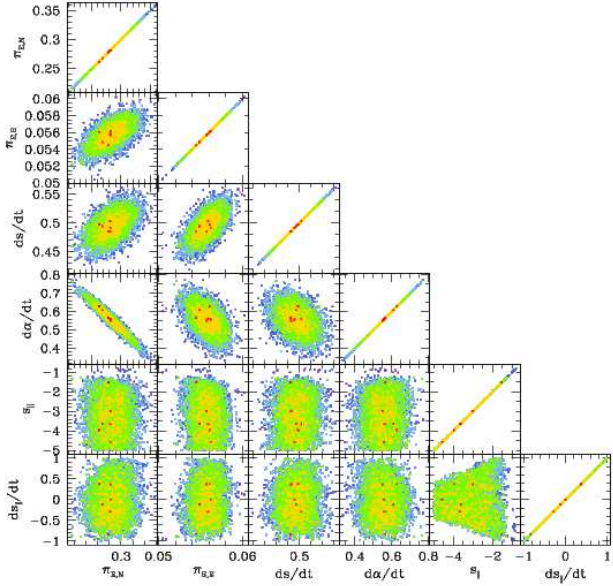}
%\epsscale{0.75}
%\plotone{lc.eps}
\caption{
$\Delta\chi^2$ distributions of MCMC points in the planes of the higher-order
lensing-parameter combinations obtained from modeling considering full
orbital parameters. The color coding indicates points 
within $1\sigma$ (red), $2\sigma$ (yellow), $3\sigma$ (green), $4\sigma$ (cyan),
and $5\sigma$ (blue).
\vskip0.4cm
}
\label{fig:five}
\end{figure}
% --------------------------------------------------------------

We note that OGLE-2018-BLG-0022 is a very rare case in which all the lensing
parameters including those describing the higher-order effects are accurately determined
without any ambiguity. As mentioned, the event does not suffer from the close/wide
degeneracy, and thus there is no ambiguity in the binary separation. Furthermore, the
ecliptic degeneracy is resolved with a significant confidence level, and thus the microlens
parallax and the resulting lens mass are uniquely determined.

Prompted by the accuracy of modeling, we further check whether the determinations of the
complete orbital parameters are possible for this event. This requires two additional
parameters $s_\parallel$ and $ds_\parallel/dt$, which represent the line-of-sight binary 
separation normalized to $\thetae$ and the change rate of $s_\parallel$, respectively 
\citep{Skowron2011, Shin2012}. One also needs the information of the angular Einstein 
radius, and we describe the procedure for the $\thetae$ estimation in section~\ref{sec:three-three}.  
We find that it is difficult to determine the full orbital parameters. In Figure~\ref{fig:five}, 
we present the $\Delta\chi^2$ distributions of MCMC points in the planes of the higher-order 
lensing-parameter combinations. It is found that the additional two parameters ($s_\parallel$ 
and $ds_\parallel/dt$) are poorly constrained, while the other higher-order parameters 
($\pien$, $\piee$, $ds/dt$, $d\alpha/dt$) are well constrained. We judge that the difficulty 
of full characterization of the orbital lens motion is caused by the short duration of the major 
anomaly features.

% Figure 6 ------------------------------------------------------
\begin{figure}
\includegraphics[width=\columnwidth]{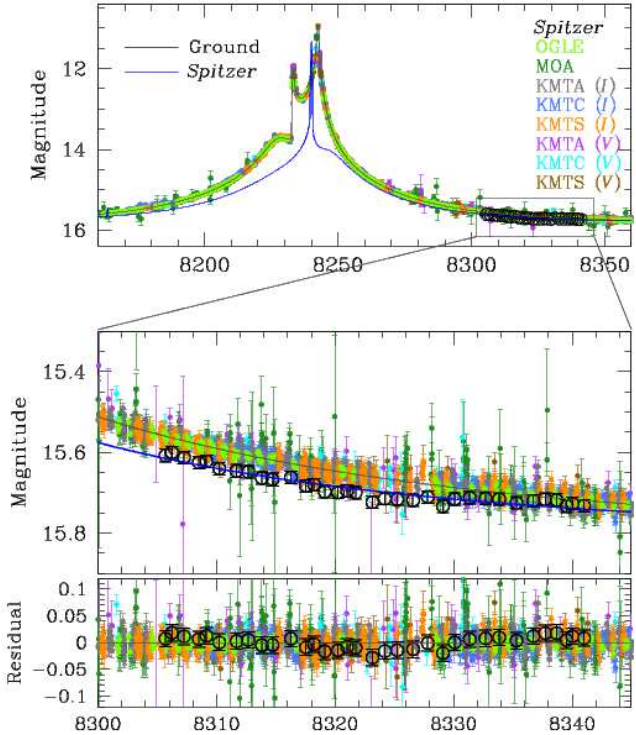}
%\epsscale{0.75}
%\plotone{lc.eps}
\caption{
Model light curves obtained from the combined analysis of the ground-based and space-based 
{\it Spitzer} data.  The black and blue curves are for the ground-based and {\it Spitzer} 
data, respectively.  The lower panels show the zoom of the region of the {\it Spitzer} data 
obtained during the period $8305.7 \leq {\rm HJD}^\prime \leq 8341.0$.
\vskip0.4cm
}
\label{fig:six}
\end{figure}
% --------------------------------------------------------------

\subsection{{\it Spitzer} Data}\label{sec:three-two}

In addition to the ground-based data, there exist data obtained from space-based 
{\it Spitzer} observations.  {\it Spitzer} observations of the event were conducted 
in 3.6~$\mu$m channel ($L$ band) with 1-day cadence during the period 
$8305.7 \leq {\rm HJD}^\prime \leq 8341.0$ and, in total, 34 data points were acquired.  
Data reduction was conducted using the procedure described in \citet{Calchi2015}.  
In Figure~\ref{fig:six}, we plot the {\it Spitzer} data over the data points from 
the ground-based observations.

{\it Spitzer} data may help to improve the accuracy of $\pie$ measurement.  This is 
because the {\it Spitzer} telescope is in a heliocentric orbit and thus the 
Earth-satellite separation and the physical Einstein radius, $r_{\rm E}=D_{\rm L}\thetae$, 
are of the same order of au.  In this case, the light curve seen from space would be 
substantially different from the light curve observed from the ground 
\citep{Refsdal1966, Gould1994}, and thus space-based data can give an important 
constraint on the microlens parallax \citep{Han2018a}.  We, therefore, check the 
effect of the {\it Spitzer} data on the $\pie$ measurement.  In the analysis with 
the additional {\it Spitzer} data, we impose a constraint of the source color with 
the measured instrumental value of $I-L = 5.21 \pm 0.02$ following the procedure 
described in \citet{Shin2017}.

% Figure 7 ------------------------------------------------------
\begin{figure}
\includegraphics[width=\columnwidth]{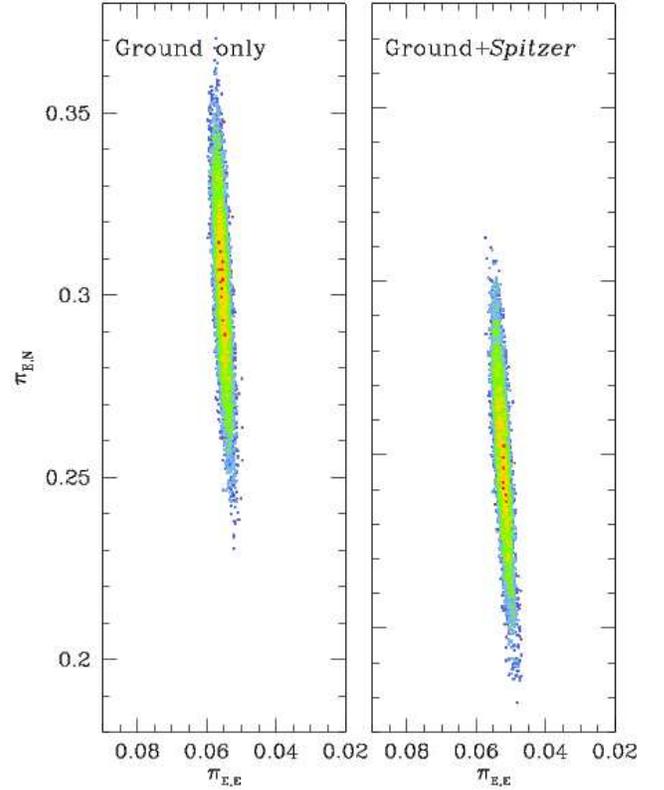}
%\epsscale{0.75}
%\plotone{lc.eps}
\caption{
$\Delta\chi^2$ distributions of MCMC points in the $\piee$--$\pien$ plane obtained 
from modelings with only ground-based data (left panel) and with additional 
space-based {\it Spitzer} data.  The color coding indicates MCMC points within 
1$\sigma$ (red), 2$\sigma$ (yellow), 3$\sigma$ (green), 4$\sigma$ (cyan), and 
5$\sigma$ (blue) levels. 
\vskip0.4cm
}
\label{fig:seven}
\end{figure}
% --------------------------------------------------------------

In the right column of Table~\ref{table:two}, we present the lensing parameters 
estimated with the additional {\it Spitzer} data.  From the comparison of the parameters 
with those estimated from the ground-based data, it is found that the parameters are 
similar to each other except for the slight differences in the higher-order parameters.  
In Figure~\ref{fig:seven}, we present the $\Delta\chi^2$ distributions of MCMC points 
in the $\piee$--$\pien$ plane obtained from the modelings with (right panel) and without 
(left panel) the {\it Spitzer} data.  It is found that the additional {\it Spitzer} 
data make the north component of the microlens parallax vector slightly smaller than 
the value estimated from the ground-based data.\\ In Figure~\ref{fig:four}, we present 
the source trajectory seen from the satellite (blue curve with an arrow marked by 
`{\it Spitzer}').  In Figure~\ref{fig:six}, we also present the model light curve 
for the {\it Spitzer} data (blue curve).

\subsection{Angular Einstein Radius}\label{sec:three-three}

We determine the angular Einstein radius, which is the other ingredient needed for the 
lens mass measurement besides $\pie$, from the combination of the normalized source radius
and the angular source radius, i.e., $\thetae=\theta_*/\rho$. The normalized source 
radius $\rho$ is determined by analyzing the caustic-crossing parts of the light curve. 
The angular source radius $\theta_*$ is estimated based on the de-reddened color, 
$(V-I)_0$, and brightness, $I_0$, of the source.  We determine $(V-I)_0$ and $I_0$
using the method of \citet{Yoo2004}, which utilizes the centroid of the red giant clump 
(RGC) in the color-magnitude diagram (CMD) as a reference.

% Figure 8 ------------------------------------------------------
\begin{figure}
\includegraphics[width=\columnwidth]{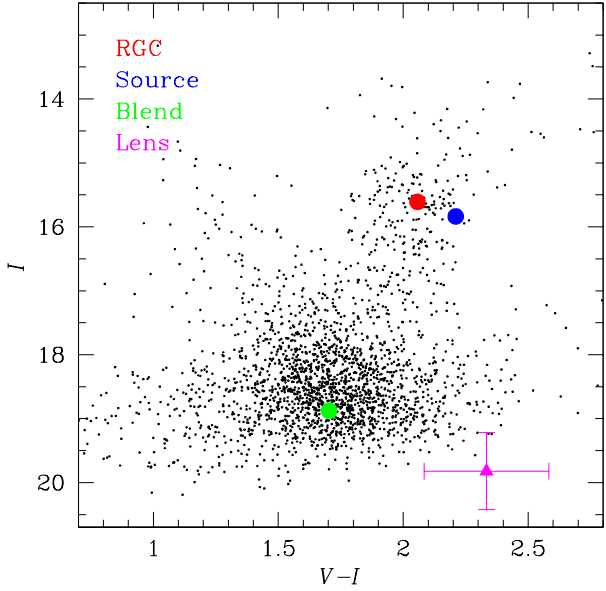}
%\epsscale{0.75}
%\plotone{lc.eps}
\caption{
Source position (blue dot) in the color-magnitude diagram of stars around 
the source.  The red dot indicates the centroid of the red giant (RGC).
The green dot represents the position of the blend, and the triangle dot with 
error bars denotes the expected position of the lens estimated from its
mass and distance.
%\vskip0.3cm
}
\label{fig:eight}
\end{figure}
% --------------------------------------------------------------

In Figure~\ref{fig:eight}, we mark the position of the source with respect to 
the RGC centroid in the OGLE-III CMD.
We note that the de-reddened source color and brightness are estimated using the 
$I$ and $V$-band pyDIA photometry of the KMTC data set.  However, the magnitude 
of the KMTC data is not calibrated, while the OGLE-III data are calibrated
\citep{Szymanski2011}.
We, therefore, place the source position on the calibrated OGLE-III CMD using the offsets 
in color and brightness between the RGC centroids of the KMTC and OGLE-III CMDs.  
With the apparent color and brightness of the source of $(V-I, I)=(2.21, 15.86)$ 
and the RGC centroid of $(V-I, I)_{\rm RGC}=(2.06, 15.61)$ combined with the known 
de-reddened values of the RGC centroid $(V-I, I)_{\rm RGC,0}=(1.06, 14.35)$ 
\citep{Bensby2011, Nataf2013}, we estimate that the de-reddened color and brightness 
of the source are $(V-I, I)_0=(1.21, 14.58)$, indicating that the 
source is a K-type giant.  The measured $V-I$ color is converted into $V-K$ color 
using the color-color relation of \citet{Bessell1988}.  We then estimate the angular 
source radius using the $(V-K)/\theta_*$ relation of \citet{Kervella2004}.  We 
estimate that the source has an angular radius of $\theta_* = 6.54 \pm 0.46~\mu{\rm as}$.

With the source radius, we estimate the angular Einstein radius of
\begin{equation}
\thetae = 1.31 \pm 0.09~{\rm mas}.
\label{eq2}
\end{equation}
Combined with the measured event timescale $t_{\rm E}$, the relative lens-source proper 
motion as measured in the geocentric frame is estimated by
\begin{equation}
\mu_{\rm geo} = {\thetae \over t_{\rm E}}=
6.82 \pm 0.48~{\rm mas}~{\rm yr}^{-1}.
\label{eq3}
\end{equation}
In the heliocentric frame, the proper motion is
\begin{equation}
\mu_{\rm helio} = 
\left| 
%\muvec_{\rm geo} 
\mu_{\rm geo}\,{\pivec_{\rm E} \over \pie} 
+ {\bf v}_{\oplus,\perp} {\pi_{\rm rel}\over {\rm au}}
\right| = 
7.33 \pm 0.52~{\rm mas}~{\rm yr}^{-1}.
\label{eq4}
\end{equation}
Here ${\bf v}_{\oplus,\perp} (N,E)=(2.1, 18.5)~{\rm km}~{\rm s}^{-1}$ denotes the 
projected velocity of Earth at $t_0$, $\pi_{\rm rel}={\rm au}(D_{\rm L}^{-1}-D_{\rm S}^{-1})$, 
and $D_{\rm S}$ denotes the distance to the source \citep{Gould2004, Dong2009}.  In 
Table~\ref{table:three}, we summarize the angular Einstein radius and the proper motion.  
Also listed is the angle of the heliocentric lens-source relative proper motion
as measured from north toward east, 
i.e., $\phi_{\rm helio}=\tan^{-1} (\mu_{{\rm helio},E}/\mu_{{\rm helio},N})\sim 22^\circ$.

% Table 3 ------------------------------------------------
\begin{deluxetable}{lcc}
\tablecaption{Einstein radius and proper motion\label{table:three}}
\tablewidth{240pt}
%\tabletypesize{\small}
\tablehead{
\multicolumn{1}{c}{Quantity}       &
\multicolumn{1}{c}{Value}  
}
\startdata                                              
Angular Einstein radius         &   1.31 $\pm$ 0.09 mas            \\
Proper motion (geocentric)      &   6.82 $\pm$ 0.48 mas yr$^{-1}$  \\
Proper motion (heliocentric)    &   7.32 $\pm$ 0.52 mas yr$^{-1}$  \\
$\phi_{\rm helio}$              &   $21.7^\circ\pm 1.5^\circ$       
\enddata                            
\tablecomments{
The angle $\phi_{\rm helio}$ represents the angle of the heliocentric lens-source 
relative proper motion as measured from north toward east. }
\end{deluxetable}
%\vskip1.0cm
% -------------------------------------------------------------

% Table 4 ------------------------------------------------
\begin{deluxetable}{lcc}
\tablecaption{Physical Lens Parameters\label{table:four}}
\tablewidth{240pt}
%\tabletypesize{\small}
\tablehead{
\multicolumn{1}{c}{Quantity}       &
\multicolumn{1}{c}{Value}  
}
\startdata                                              
Mass of the primary lens       &  0.40  $\pm$ 0.05 $M_\odot$ \\
Mass of the companion lens     &  0.15  $\pm$ 0.01 $M_\odot$ \\
Distance to the lens           &  2.21  $\pm$ 0.18 kpc       \\
Projected separation           &  1.54  $\pm$ 0.13 au        \\
$({\rm KE}/{\rm PE})_\perp$    &  0.09  $\pm$ 0.01
\enddata                            
\tablecomments{$({\rm KE}/{\rm PE})_\perp$ represents the projected kinetic-to-potential 
energy ratio of the lens system.
\bigskip
  }
\end{deluxetable}
%\vskip0.5cm

% -------------------------------------------------------------

\subsection{Physical Lens Parameters}\label{sec:three-four}

Being able to determine $\pie$ and $\thetae$ without any ambiguity, the mass of the lens 
is uniquely determined. It is found that the lens is a binary composed of an early M-dwarf
primary with a mass
\begin{equation}
M_1 = 0.50 \pm 0.05~M_\odot
\label{eq5}
\end{equation}
and a late M-dwarf companion with a mass
\begin{equation}
M_2 = 0.15 \pm 0.01~M_\odot.
\label{eq6}
\end{equation}
We note that the masses are estimated based on the solution obtained using both the 
ground-based and {\it Spitzer} data.

With the determined $\pie$ and $\thetae$, the distance to the lens is determined by
\begin{equation}
D_{\rm L} = {{\rm au} \over \pie\thetae + \pi_{\rm S} } = 
2.21 \pm 0.18~{\rm  kpc},
\label{eq7}
\end{equation}
indicating that the lens is in the disk.  Here $\pi_{\rm S}={\rm au}/D_{\rm S}$. 
The source distance is estimated using the relation $D_{\rm S}=d_{\rm GC}/(\cos l 
+ \sin l \cos\theta_{\rm bar}/\sin\theta_{\rm bar})\sim 7.87~{\rm kpc}$, where 
$d_{\rm GC}\sim 8160$ pc is the distance to the Galactic center and 
$\theta_{\rm bar}=40^\circ$ is the orientation angle of the bulge bar 
\citep{Nataf2013}. Once the distance is estimated, the projected separation between 
the lens components is estimated by
\begin{equation}
a_\perp = sD_{\rm L}\thetae = 
1.54 \pm 0.13~{\rm au}.
\label{eq8}
\end{equation}

For the source of the event, the parallax is not measured but the proper motion (in 
the heliocentric frame) is listed in the {\it Gaia} archive\footnote{https://archives.esac.esa.int/gaia}
with values
\begin{equation}
\muvec_{\rm S}(N,E) = (-6.12, -2.76)~{\rm mas}~{\rm yr}^{-1}.
\label{eq9}
\end{equation}
The proper motion indicates that the source is a typical bulge star.
Since the relative lens-source proper motion (in the heliocentric frame), $\muvec$, is 
related to the proper motions of the source, $\muvec_{\rm S}$, and the lens, $\muvec_{\rm L}$, 
by $\muvec=\muvec_{\rm L}-\muvec_{\rm S}$, the {\it Gaia} measurement of the source proper motion 
allows us to estimate the lens proper motion by the relation 
\begin{equation}
\muvec_{\rm L}(N,E) = \muvec_{\rm S} + \muvec = 
(0.69, -0.05)~{\rm mas}~{\rm yr}^{-1}.
\label{eq10}
\end{equation}
Then, the projected lens velocity in the heliocentric frame is 
\begin{equation}
{\bf v}_{{\rm L},\perp}(N,E)=\muvec_{\rm L} D_{\rm L}= 
(7.2, -0.5)~{\rm km}~{\rm s}^{-1},
\label{eq11}
\end{equation}
which is very typical for disk stars. Therefore, the estimated lens distance is 
consistent with the additional constraint from the {\it Gaia} observation.

% Figure 9 ------------------------------------------------------
\begin{figure}
\includegraphics[width=\columnwidth]{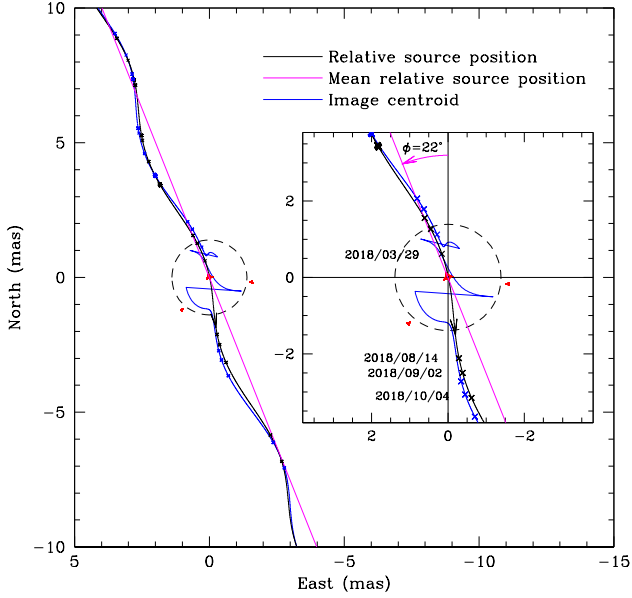}
%\epsscale{0.75}
%\plotone{lc.eps}
\caption{
Motions of the source (black curve) and image centroid (blue curve) predicted
by the photometric lensing solution. The straight magenta line represents the relative
lens-source motion. The points on the source and image-centroid curves represent the
times of Gaia observations. Coordinates are centered at the barycenter of the lens and
the abscissa and ordinate are aligned to the East and North directions.
%\vskip0.5cm
}
\label{fig:nine}
\end{figure}
% --------------------------------------------------------------

We also check the validity of the lensing solution by computing the projected
kinetic-to-potential energy ratio of the lens system. 
From the determined physical parameters $M$ and $a_\perp$ combined with the 
lensing parameters $s$, $ds/dt$, and $d\alpha/dt$, 
the ratio is computed by
\begin{equation}
\left( {{\rm KE} \over {\rm PE}}\right)_\perp
={(a_\perp/{\rm au})^3\over 8\pi^2(M/M_\odot) }
\left[
\left( {1\over s}{ds/dt \over {\rm yr}^{-1}}\right)^2 +
\left( {d\alpha/dt\over {\rm yr}^{-1}} \right)^2
\right]
\sim 0.09.
\label{eq12}
\end{equation}
In order for the lens to be a gravitationally bound system, the ratio should meet the 
condition of $({\rm KE}/{\rm PE})_\perp \leq {\rm KE}/{\rm PE} \leq 1.0$. 
The estimated kinetic-to-potential energy ratio satisfies this condition.
In Table~\ref{table:four}, we summarize the determined physical lens parameters.

% Figure 10 ------------------------------------------------------
\begin{figure}
\includegraphics[width=\columnwidth]{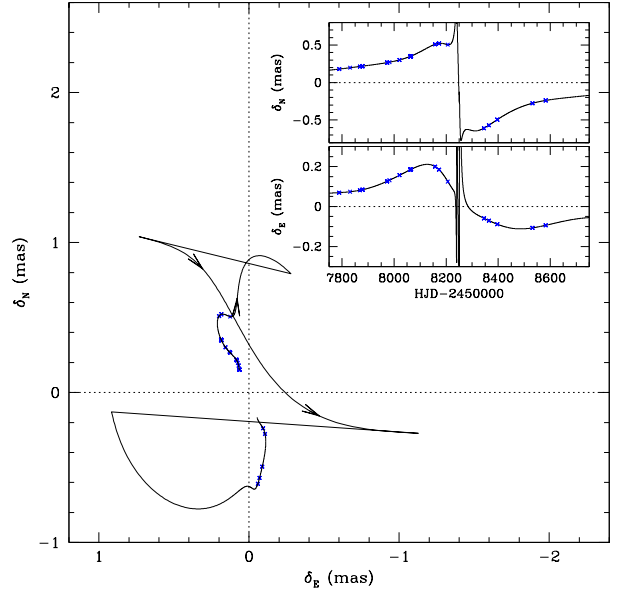}
%\epsscale{0.75}
%\plotone{lc.eps}
\caption{
Shift of the image centroid with respect to the unlensed source position. The
two insets show the north and east components of the centroid shift vector as a
function of time.
\vskip0.5cm
}
\label{fig:ten}
\end{figure}
% --------------------------------------------------------------

Given that the distance to the lens is small, the lens might comprise an important 
fraction of the blended light, e.g., OGLE-2017-BLG-0039 \citep{Han2018b}.  We check 
this possibility by inspecting the agreement between the positions of the lens and 
blend in the CMD.  In Figure~\ref{fig:eight}, we place the locations of the blend 
and lens.  The lens location is estimated based on the mass and distance.  Considering 
the close distance to the lens, we assume that the lens experiences $\sim 1/3$ of the 
total extinction and reddening toward the bulge field of $A_I\sim 1.26$ and $E(V-I)\sim 1.02$, 
respectively  \citep{Nataf2013}.  The estimated color and brightness of the lens are 
$(V-I,I)_{\rm L}\sim (2.3, 19.8)$, while those of the blend are $(V-I,I)_{b}\sim (1.7, 18.9)$.  
The lens is substantially fainter and redder than the blend and this indicates that the 
lens is not the main source of the blended light.

\section{Prediction of {\it Gaia} Astrometric Measurements}\label{sec:four}

A lensing phenomenon causes not only the magnification of the source brightness but also
the change of the image positions. When a source star is gravitationally lensed, it
is split into multiple images and the brightness and location of each image change 
with the change of the relative lens-source position. 
By employing the GRAVITY instrument of the  Very Large Telescope Interferometer (VLTI), 
\cite{Dong2018} recently reported the first resolution of the two microlens images 
of a domestic microlensing event TCP J05074264+2447555 \citep{Gaia2018},  which occurred 
on a nearby source star located within $\sim 1$--2 kpc of the Sun in the direction opposite 
to the bulge field.  This demonstrates that resolving microlens images is possible provided 
that a source is bright enough for VLTI GRAVITY observations ($K\lesssim 10.5$ or 
$I\lesssim 12.8$ for an $I-K=2.3$ star).  For OGLE-2018-BLG-0022, the event during the 
caustic crossings was brighter than this threshold magnitude, and thus the separate 
images could have been resolved from VLTI GRAVITY observations, but no such observation 
was conducted.

Without directly resolving the separate images, a binary lens can still be astrometrically 
constrained by measuring the positional displacement of the image centroid \citep{Han2001}.
Compared to the direct image resolution, which requires a resolution of order mas, 
the centroid shift measurement can be done with {\it Gaia}, which has resolution 
lower than VLTI GRAVITY by two orders of magnitude.  Actually, the source of the event 
OGLE-2018-BLG-0022 is in the second {\it Gaia} data release.  We, therefore, predict the 
astrometric behavior of the image centroid motion based on the solution obtained from 
the analysis of the photometric data.

In Figure~\ref{fig:nine}, we present the motions of the source (black curve) and the
image centroid (blue curve) in the East-North coordinates. By the time of writing this
paper, the field has been observed 73 times by {\it Gaia} since 2014/10/15 
(${\rm HJD}^\prime\sim 6945$) and 4 forthcoming observations are scheduled until 
2019/04/08 (${\rm HJD}^\prime\sim 8583$). We mark the positions of the {\it Gaia} 
observations on the curves of the source and image-centroid motions. The straight 
magenta line represents the mean relative lens-source proper motion, i.e., without parallax 
motion, which is heading toward south-west with an angle 
$\phi_{\rm helio} \simeq 22^\circ$ as measured North through East.
In Figure~\ref{fig:ten}, we present the shift of the image 
centroid with respect to the unlensed source position, $\deltavec$. In the two insets, 
we present the north and east components of $\deltavec$ as a function of time.

The expected signal-to-noise ratio of the astrometric centroid shift measurement is 
\begin{equation}
{S\over N} =
{\sqrt{\sum_{i=1}^{N_{\rm obs}}{\delta_i^2}}\over \sqrt{2}\sigma_{\rm 1D}}
= {3.04\,{\rm mas}\over 0.707 {\rm mas}} = 4.3,
\label{eq13}
\end{equation}
where $\delta_i = (\delta_{N,i}^2 + \delta_{E,i}^2)^{1/2}$ is the
astrometric deviation, and $\sigma_{\rm 1D}$ is the mean astrometric
error of an individual {\it Gaia} measurement along its principle
axis.  We estimate
$\sqrt{2}\sigma_{\rm 1D}= \sigma_\pi
[ N_{\rm meas}\langle\sin^2(\Delta l)_{\rm eclip}\rangle]^{1/2}$,
where $\sigma_\pi = 0.218\,{\rm mas}$ is the reported {\rm Gaia}
parallax error, $N_{\rm meas}=13$ is the number of {\it Gaia} epochs
entering this measurement, and
$[\langle\sin^2(\Delta l)_{\rm eclip}\rangle]^{1/2}=0.9$ is the
RMS parallactic offset of the target as seen by {\it Gaia} at the
times of the observations.
Therefore, it is expected that the astrometric centroid shift 
can be reliably measured.

\section{Conclusion}\label{sec:five}

We analyzed the binary-lensing event OGLE-2018-BLG-0022.  Thanks to the dense and 
continuous coverage with the high-quality photometry data from the ground-based 
observations combined with space-based {\it Spitzer} observations, we found a lensing 
solution including microlens-parallax and lens-orbital parameters without any ambiguity, 
leading to the unique determination of the physical lens parameters.  The robust and 
unique solution and the relatively bright source enabled the prediction of astrometric 
lensing signals that could be confirmed from actual astrometric observations using 
{\it Gaia}.

\acknowledgments
Work by CH was supported by the grant (2017R1A4A1015178) of National Research Foundation of Korea.
% Gould  
Work by AG was supported by US NSF grant AST-1516842.
Work by IGS and AG were supported by JPL grant 1500811.
AG received support from the
European Research Council under the European Union's
Seventh Framework Programme (FP 7) ERC Grant Agreement n.~[321035].
%MOA 
The MOA project is supported by JSPS KAKENHI Grant Number JSPS24253004,
JSPS26247023, JSPS23340064, JSPS15H00781, and JP16H06287.
YM acknowledges the support by the grant JP14002006.
DPB, AB, and CR were supported by NASA through grant NASA-80NSSC18K0274. 
The work by CR was supported by an appointment to the NASA Postdoctoral Program at the Goddard 
Space Flight Center, administered by USRA through a contract with NASA. NJR is a Royal Society 
of New Zealand Rutherford Discovery Fellow.
% OGLE  
The OGLE project has received funding from the National Science Centre, Poland, grant
MAESTRO 2014/14/A/ST9/00121 to AU.
% KMTNet
This research has made use of the KMTNet system operated by the Korea
Astronomy and Space Science Institute (KASI) and the data were obtained at
three host sites of CTIO in Chile, SAAO in South Africa, and SSO in
Australia.
% Spitzer 
This work is based (in part) on observations made with the $Spitzer$ Space
Telescope, which is operated by the Jet Propulsion Laboratory, California
Institute of Technology under a contract with NASA. Support for this work
was provided by NASA through an award issued by JPL/Caltech.
% KREONET
We acknowledge the high-speed internet service (KREONET)
provided by Korea Institute of Science and Technology Information (KISTI).


\begin{thebibliography}{}

\bibitem[Alard \& Lupton(1998)]{Alard1998} Alard, C., \& Lupton, R.~H.\ 1998, \apj, 503, 325
\bibitem[Albrow(2017)]{Albrow2017} Albrow, M.\ 2017, MichaelDAlbrow/pyDIA: Initial Release on Github, doi: 10.5281/zenodo.268049
\bibitem[Albrow et al.(2009)]{Albrow2009} Albrow, M.~D., Horne, K., Bramich, D.~M., et al.\ 2009, \mnras, 397, 2099
\bibitem[An(2005)]{An2005} An, J.~H.\ 2005, \mnras, 356, 1409
\bibitem[Batista et al.(2011)]{Batista2011} Batista, V., Gould, A., Dieters, S., et al.\ 2011, \aap, 529, 102
\bibitem[Bensby et al.(2011)]{Bensby2011} Bensby, T., Ad\'en, D., Mel\'endez, J., et al.\ 2011, \pasp, 533, 134
\bibitem[Bessell \& Brett(1988)]{Bessell1988} Bessell, M.~S., \& Brett, J.~M. 1988, \pasp, 100, 1134
\bibitem[Bond et al.(2001)]{Bond2001} Bond, I. A., Abe, F., Dodd, R. J., et al. 2001, \mnras, 327, 868
\bibitem[Bramich(2018)]{Bramich2018a} Bramich, D.~M.\ 2018, \aap, 618, A44
\bibitem[Bramich \& Nielsen(2018)]{Bramich2018b} Bramich, D.~M., \& Nielsen, M.~B.\ 2018, Acta Astron., 68, 183
\bibitem[Calchi Novati et al.(2015)]{Calchi2015} Calchi Novati, S., Gould, A., Yee, J.~C., 2015, \apj, 814, 92
\bibitem[Claret(2000)]{Claret2000} Claret, A.\ 2000, \aap, 363, 1081
\bibitem[Dominik(1998)]{Dominik1998} Dominik, M.\ 1998, \aap, 329, 36
\bibitem[Dominik(1999)]{Dominik1999} Dominik, M.\ 1999, \aap, 349, 108
\bibitem[Dominik et al.(2008)]{Dominik2008} Dominik, M., Horne, K., Allan, A., et al.\ 2008, Astronomische Nachrichten, 329, 248
\bibitem[Dong et al.(2009)]{Dong2009} Dong, S., Gould, A., Udalski, A., et al.\ 2009, \apj, 695, 970
\bibitem[Dong et al.(2018)]{Dong2018} Dong, S., M\'erand, A., Delplancke-Str\"obele, F., et al.\ 2018, arXiv:1809.08243
\bibitem[{\it Gaia} Collaboration(2018)]{Gaia2018} {\it Gaia} Collaboration: Brown, A.~G.~A., Vallenari, A., Prusti, T., et al.\ 2018, \aap, 616, A1
\bibitem[{\it Gaia} Collaboration(2016)]{Gaia2016} {\it Gaia} Collaboration, 2016, \aap, 595, A1
\bibitem[Gould(1992)]{Gould1992} Gould, A.\ 1992, \apj, 392, 442
\bibitem[Gould(1994)]{Gould1994} Gould, A.\ 1994, \apjl, 421, L75
\bibitem[Gould(2004)]{Gould2004} Gould, A.\ 2004, \apjl, 606, 313
\bibitem[Griest \& Safazadeh(1998)]{Griest1998} Griest, K., \& Safazadeh, N.\ 1998, \apj, 500, 37
\bibitem[Han(2001)]{Han2001} Han, C.\ 2001, \mnras, 325, 1281
\bibitem[Han et al.(2018a)]{Han2018a} Han, C., Calchi Novati, S., Udalski, A., et al.\ 2018, \apj, 859, 82
\bibitem[Han et al.(2018b)]{Han2018b} Han, C., Jung, Y. K., Udalski, A., et al.\ 2018, \apj, 867, 136
\bibitem[Han et al.(2016)]{Han2016} Han, C., Udalski, A., Lee, C.-U., et al.\ 2016, \apj, 827, 11
\bibitem[Ioka et al.(1999)]{Ioka1999} Ioka, K., Nishi, R., \& Kan-Ya, Y.\ 1999, PThPh, 102, 98
\bibitem[Jung et al.(2018)]{Jung2018} Jung, Y. K., Han, C., Udalski, A., et al.\ 2018, \apj, 863, 22
\bibitem[Kervella et al.(2004)]{Kervella2004} Kervella, P., Th\'evenin, F., Di Folco, E., \& S\'egransan, D.\ 2004, \aap, 426, 29
\bibitem[Kim et al.(2016)]{Kim2016} Kim, S.-L., Lee, C.-U., Park, B.-G., et al.\ 2016, JKAS, 49, 37
\bibitem[Nataf et al.(2013)]{Nataf2013} Nataf, D.~M., Gould, A., Fouqu\'e, P., et al.\ 2013, \apj, 769, 88
\bibitem[Paczy\'nski(1996)]{Paczynski1996} Paczy\'nski, B.\ 1996, Acta Astron., 46, 291
\bibitem[Proft et al.(2011)]{Proft2011} Proft, S., Demleitner, M., \& Wambsganss, J.\ 2011, \aap, 536, 50 
\bibitem[Refsdal(1966)]{Refsdal1966} Refsdal, S.\ 1966, \mnras, 134, 315
\bibitem[Sahu et al.(2014)]{Sahu2014} Sahu, K., Bond, H.~E., Anderson, J., \& Dominik, M.\ 2014, \apj, 782, 89
\bibitem[Salim \& Gould(2000)]{Salim2000} Salim, S, \& Gould, A,\ 2000, \apj, 539, 241
\bibitem[Schneider \& Weiss(1986)]{Schneider1986} Schneider, P., \& Weiss, A.\ 1986, \aap, 164, 237
\bibitem[Shin et al.(2012)]{Shin2012} Shin, I.-G., Han, C., Choi, J.-Y., et al.\ 2012, \apj, 755, 91
\bibitem[Shin et al.(2017)]{Shin2017} Shin, I.-G., Udalski, A., Yee, J. C., et al. 2017, \aj, 154, 176
\bibitem[Skowron et al.(2011)]{Skowron2011}Skowron, J., Udalski, A., Gould, A., et al.\ 2011, \apj, 738, 87
\bibitem[Smith et al.(2003)]{Smith2003} Smith, M.~C., Mao, S., \& Paczy\'nski, B. 2003, \mnras, 339, 925
\bibitem[Szyma\'nski et al.(2011)]{Szymanski2011}Szyma\'nski, M.K., Udalski, A., Soszy\'nski, I., et al. 2011, Acta Astron., 61, 83
\bibitem[Udalski(2003)]{Udalski2003} Udalski, A.\ 2003, Acta Astron., 53, 291
\bibitem[Udalski et al.(2015)]{Udalski2015} Udalski,~A., Szyma\'nski,~M.~K., \& Szyma\'nski,~G.\ 2015, Acta Astron., 65, 1
\bibitem[Yoo et al.(2004)]{Yoo2004} Yoo, J., DePoy, D.~L., Gal-Yam, A., et al.\ 2004, \apj, 603, 139
\bibitem[Zhu et al.(2015)]{Zhu2015} Zhu, W., Udalski, A., Gould, A., et al.\ 2015, \apj, 805, 8

\end{thebibliography}
\end{document}